\documentclass[11pt]{article}
\usepackage{moriond,epsfig}

\def\be{\begin{equation}}
\def\ee{\end{equation}}
\def\bea{\begin{eqnarray}}
\def\eea{\end{eqnarray}}

\begin{document}
\vspace*{4cm}
\title{CLEO-c and CESR-c: A New Frontier of Electroweak and QCD Physics}

\author{ K. Benslama \\
	for the CLEO Collaboration}

\address{University of Illinois, Loomis Lab, 1110 W. Green St.,\\
Urbana, IL 61801 USA}

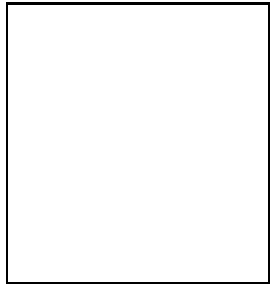
\maketitle\abstracts{ The new proposed experiment CLEO-c in the Wilson Laboratory at Cornell University will explore the physics potential of the CLEO detector and the CESR storage ring operation in the center-of-mass energy range 3 - 5 GeV. Data taking could start as early as January 2003. Estimates of the luminosity that a suitably modified CESR can deliver in this energy range imply data samples of 1 - 4 $fb^{-1}/year$. The physics program of CLEO-c can be divided into two parts: weak interaction physics and QCD physics. The electroweak program presented at the XXXVIIth Rencontres de Moriond is discussed.
}

\section{Introduction and Motivation}

The CLEO-c Physics program includes a set of measurements that will substantially advance our understanding of important Standard Model processes and set the stage for understanding the larger theory in which we imagine the Standard Model to be embedded. Moreover the CLEO-c program also include searches for physics beyond the Standard Model.

In this note we will describe the CLEO-c measurements that will probe the essential nature of the weak decays of charm mesons and flavor mixing by clean, high-statistics studies of semileptonic charm decays and $D\bar{D}$ mixing. We will describe precision measurements of branching fractions that will set the absolute scale and validate the theoretical techniques for much of the heavy flavor physics done in the next decade, and will describe direct searches for physics within and beyond the Standard Model, searches that are unique to the CLEO-c program.

\section{Run Plan and Datasets}\label{runplan}

The CESR accelerator will operate at the center-of-mass energies corresponding to $\sqrt{s} \sim 4140, \;\sqrt{s} \sim 3770(\psi^{"}), \;and \;\sqrt{s} \sim 3100(J/\psi)$ for approximately one calendar year each. Taking into account the anticipated luminosity which will range from $5 \times 10^{32} cm^{-2} s^{-1}$ down to about $1 \times 10^{32} cm^{-2} s^{-1}$ over this energy range, the proposed run plan will yield 3 $fb^{-1}$ each at the end $\psi^{"}$ and at $\sqrt{s} \sim 4140$ above $D_{s}\bar{D_{s}}$ threshold, and 1 $fb^{-1}$ at the $J/\psi$. These integrated luminosities correspond to samples of 1.5 million $D_{s}\bar{D_{s}}$ pairs, 30 million $D\bar{D}$ pairs, and one billion $\psi$ decays. As a point of reference , we note that these datasets will exceed those of the MARK III experiment by factors of 480, 310, and 170, respectively. If time permits, we will take additional data samples at the $\Lambda_{c}\bar{\Lambda_{c}}$ threshold region, the $\tau^{+}\tau^{-}$ threshold region, and the $\psi(3684)$.

In addition, prior to the conversion to low energy operation, we will take $\sim$ 4 $fb^{-1}$ spread over the $\Upsilon(1S), \Upsilon(2S), \Upsilon(3S)$ resonances to launch the QCD part of the program. These datasets will increase the available $b\bar{b}$ bound state data by more than an order of magnitude.

\section{Measurement}

Our principal measurement targets include:
\subsection{Leptonic charm decays:}
$D^{-} \rightarrow l^{-}\bar{\nu} \;and \;D^{-}_{s} \rightarrow l^{-}\bar{\nu} $
From the muonic decays alone we can determine the decay constants $f_{D}$ and $f_{D_{s}}$ to a precision of about $2\%$. We note that $f_{D}$ and $f_{D_{s}}$ are only known to about 35$\%$ and 100$\%$ respectively, and $f_{B}$ and $f_{B_{s}}$ are unlikely to be measured to any useful precision in this decade.

\subsection{Semileptonic charm decays:}
$ D \rightarrow (K, K^{*})l\nu, D \rightarrow (\pi,\rho,\omega)l\nu, D_{s} \rightarrow (\eta,\phi)l\nu, D_{s} \rightarrow (K, K^{*})l\nu, \;and \;\Lambda_{c} \rightarrow \Lambda l \nu$.
Absolute branching rations in critically interesting modes like $D \rightarrow \pi l \nu \; and \; D \rightarrow K l \nu$ will be measured to $1\%$, and form factors slopes to $4\%$. Form factors in all modes can be measured across the full range of $q^{2}$ with excellent resolution. Semileptonic decays are the primary source of data for the CKM elements $|V_{ub}|, |V_{cb}|, |V_{cd}|,\;and \;|V_{cs}|$, but these CKM elements can not be extracted without accurate knowledge of the form factors. Currently, semileptonic branching rations are known with uncertainties that range from $5\%$ to $73\%$. Inclusive semileptonic decays such as $	D \rightarrow e X$, $D_{s} \rightarrow e X$, and $\Lambda_{c} \rightarrow e X$ will also be examined and branching rations will be measured to a precision of 1 - 5 $\%$. Currently, such quantities are known with uncertainties that range from 4$\%$ to 63$\%$.
\subsection{Search for Hadronic decays of charmed mesons.}
We will establish the critical normalizing modes $D \rightarrow K\pi$, $D^{+} \rightarrow K\pi\pi$, and $D_{s} \rightarrow \phi \pi$ to a precision of order 1 - 2 $\%$. Currently these are known with uncertainties that range from up to 25$\%$ and are even larger for other hadronic decays of interest. Many important B meson branching rations are normalized with respect to these charm modes.
\subsection{Rare decays, $D\bar{D}$ mixing, and CP violating decays}
We will search rare decays with a typical sensitivity of $10^{-6}$, study mixing with a sensitivity to $ x = \frac{\Delta M}{M} \; and \; y = \frac{\Delta \Gamma}{\Gamma}$ of under $1\%$, and detect CP violating asymmetries that may be present with a sensitivity of better than $1\%$. We will also search $\tau$ decays for evidence of new physics.
\subsection{Quarkonia and QCD}
With the Largest $J/\psi$ data sample in the World, the CLEO-c QCD program will be able to determine the composition for a variety of the exotica in the mass region accessible from $J/\psi$ radiative decays. We will explore the region of $1 < M_{X} < 3 \;GeV/c^{2}$ with partial wave analyses for evidence of scalar or tensor glueballs, glueball-$q\bar{q}$ mixtures, exotic quantum numbers, quark-glue hybrids, and other evidence for new forms of matter predicted by QCD but not yet cleanly observed.
\section{Unique Features of the CLEO-c Program}
Many measurement described above have been done or attempted by other experiments such as Mark III and BES, and many are accessible to B-factory experiments operating at the $\Upsilon(4s)$. What makes CLEO-c unique ?

Compared to Mark III and BES experiments which have taken data on the same $\psi$ resonances as we discussed here, CLEO-c will have:
\subsection{Vastly more data.}
As noted above the CLEO-c data sample will be $\sim$ 200 - 500 times larger than the corresponding Mark III datasets and we will have 270 times as much D and $D_{s}$ data, and 20 times as many $\psi(3100)$ decays.
\subsection{A modern detector.}
CLEOIII is superior to both BES and Mark III detectors by substantial margins. Photon energy, for example, is factors of 10-20 times better; charger particle momentum resolution is 2-3 times better. Particle identification with the RICH and dE/dx gives tens to hundreds of sigma $K\pi$ separation across the full kinematic range.

On the other hand, CLEO-c will not have any advantage in statistics or in detector performance when compared to Babar and Belle. With an anticipated 400 $fb^{-1}$ of $\Upsilon(4s)$ data, Babar and Belle will each have about 500 million continuum $e^{+}e^{-} \rightarrow c\bar{c}$ events. Yet the data CLEO-c takes at charm threshold has distinct and powerful advantages over continuum charm production data taken at B-factories, which we list here:
\begin{itemize}
\item Charm events produced at threshold are extremely clean
\item Charm events produced at threshold are pure $D\bar{D}$
\item Double-tag studies are pristine
\item Signal/Background is optimum at threshold
\item neutrino reconstruction is clean
\item Quantum coherence: For D mixing and some CP violation studies, the fact that the D and $\bar{D}$ are produced in a coherent quantum state in $\psi(3770)$ decay is of central importance for the subsequent evolution and decay of these particles. The same is true for the $CP = +1$ mode $\psi(4140) \rightarrow \gamma D\bar{D}$. The coherence of the two initial-state particles allows simple methods to measure $D\bar{D}$ mixing parameters and check for direct CP violation.
\end{itemize}
\section{The Impact of CLEO-c}
The extensive set of 1 - 2 $\%$ precision measurements by CLEO-c will rigorously constrain theoretical calculations. The calculation which survive these tests will be validated for use in a wide variety of areas where the interesting physics cannot be extracted without theoretical input. This broader impact of CLEO-c results extends beyond the borders of CLEO-c measurement and affects most of the core issues in heavy flavor physics. We list here some of the areas that will be most notable:
\begin{itemize}
\item 5$\%$ or better precision of $|V_{ub}|$.
\item $\sim 5\%$ determinations of $|V_{td}|$ and $|V_{ts}|$
\item $\sim 1\%$ accuracy in extracting $|V_{cd}|$ and $|V_{cs}|$
\item measurement of the normalizing branching rations at the sub-percent level.\item Unitarity tests of CKM can be probed with $1\%$ precision when $|V_{cd}|$ and $|V_{cs}|$ are provided at this level by CLEO-c.
\item Over-constraining the Standard Unitarity Triangle: Provided $|V_{ub}|$ and $|V_{td}|$ have been determined at the $5\%$ level, the triangle sides will have been measured with a precision comparable to the phase quantity $\sin2\beta$, thereby allowing for the first time meaningful comparison of the sides of the unitarity triangle with one of the angles.
\item In the quarkonium studies new forms of gluonic matter may be identified.
\end{itemize}
\section{Conclusion}
The acceptance, resolution, and particle identification of the CLEO III detector are an outstanding match to the requirements of a program of precision measurements in the charm threshold region.
CLEO-c will provide precision measurements at the $1\%$ level:
\begin{itemize}
\item Of absolute reference hadronic branching fractions for D decays
\item Of $|V_{cd}|f_{D^{+}}$ and $|V_{cs}|f_{D_{s}^{+}}$ from leptonic $D^{+}$ and $D_{s}^{+}$ decays
\item Of $|V_{cd}|$ and $|V_{cs}|$ from leptonic $D^{0}$ and $D^{+}$ decays
\item With very small and well-controlled backgrounds
\item With very small and well-understood systematic errors.
\end{itemize}
These measurements will challenge the ability of LQCD to calculate decay constants and form factors for D decays, which will determine their utility for these calculations in B decays. The advances in strong-interaction calculations that we expect to drive will in turn underwrite advances in weak interaction physics not only in CLEO-c, but in all heavy-quark endeavors and in future explorations of physics beyond the Standard Model.

\end{document}